# Scrum of Scrums Solution for Large Size Teams Using Scrum Methodology

Saja Al Qurashi, M. Rizwan Jameel Qureshi

Faculty of Computing and Information Technology, King Abdulaziz University, Jeddah, Saudi Arabia
saja_alqurashi@yahoo.com, anriz@hotmail.com

**Abstract:** Scrum is a structured framework to support complex product development. However, Scrum methodology faces a challenge of managing large teams. To address this challenge, in this paper we propose a solution called Scrum of Scrums. In Scrum of Scrums, we divide the Scrum team into teams of the right size, and then organize them hierarchically into a Scrum of Scrums. The main goals of the proposed solution are to optimize communication between teams in Scrum of Scrums; to make the system work after integration of all parts; to reduce the dependencies between the parts of system; and to prevent the duplication of parts in the system.

## 1.      Introduction

The software development domain must keep pace with ongoing changes in the environment. Agile methodology is a perfect choice for any organization to produce their product in short time. Scrum is an agile methodology, which is a framework structured to support complex product development. There are three core roles in Scrum: product owner (PO), Scrum master (SM) and Scrum team (ST). The PO is a key stakeholder of the project and is responsible to visualize and prioritize the features list for the product. The SM ensures that the Scrum process is going as agreed, and prevents any impediments that can be faced by the team, e.g. communication, dependencies, etc. All the members working together to complete the set of work they have collectively committed to complete within a sprint comprise ST. The main feature of Scrum is reduced time and small team size. Organizations employ Scrum methodology to complete large projects in short time. In case small team size is inappropriate to finish a project in time, large teams are set up. The large teams in Scrum methodology may have several problems including duplication of work, communication failure, integration with other teams and dependencies between tasks in different teams (Mundra et al., 2013).

The Scrum of Scrums provides solution to these problems. The Scrum of Scrums meeting is an important technique in scaling Scrum to project where large teams are required. These meetings allow groups of teams to discuss their work, focusing on areas of overlap and integration. The attendees of the meetings should change over the course of a project. The team should choose its representative based on who will be in the best position to understand and discuss the issues that arise at that time during a project.

The Scrum of Scrums meeting is different from the regular Scrum; we describe these differences in later sections. The rest of the paper is organized as under. The related work is reviewed in Section 2. Section 3 describes the exact problem statement. Our proposed solution is described in Section 4 and 5. Section 6 describes the goals we established and a survey based on our goals to validate our proposed approach. The study is concluded in section 7.

## 2.      Related Work

Azham (2011) proposes the integration of security principles in development phases using Scrum and suggest the element of security backlog that can be used as security features analysis and implementation in Scrum phases. But the result of the proposed solution will be presented in the near future after enough data has been collected from various surveys, interviews and experiments that are underway and planned (Azham, 2011).

Chhavi et al. (2013) present Scrum to complete the work in short iterations. Automation can be beneficial at time of managing various activities of Scrum. It provides fast solution, increase reliability, repeatability, comprehensiveness and efficiently. There is a need for more research about automation of Scrum (Chhavi et al., 2013).

Noor et al. (2013) observed the progress of a project that used Scrum, through burn down charts, where remaining task is plotted against working days and actual progress line are compared to overall progress. They proposed the modified version of original burn down chart. But we must try to upgrade the chart in the future to identify more reasons of deviations (Noor et al., 2013).

Raghaw et al. (2012) present a study to identify important issues and challenges that effect quality of game development by agile method. These challenges include team management, lack of accountability, trust and confidence, documentation, work environment and training. They proposed the following guidelines: each member in the team must have knowledge and skills pertinent to the project the team members are going to work on; the project manger should not be a bottle neck to Scrum teams; new employees should be given enough time to understand the system and Scrum method; the project manager must watch if there is a lack of trust and confidence; the duration of Scrum meetings should be strictly observed; reduce documentation significantly. But the main limitation is that the project narrowed down for two firms only (Raghaw et al., 2012).

Akhtar et al. (2010) reported that Pakistan's software industry is comparatively young as compared to the global software industry. So it is flexible to adopt new project management practices and software development methodologies. The Scrum adoption, implementation and acceptance in Pakistan's software industry will be helpful for making accurate decision. But, so far the Scrum is not very popular in Pakistan's software industry (Akhtar et al., 2010).

Vlaanderen et al. (2009) present a case study of software product management based on Scrum method. They argued that the product manager can cope with complex requirements in agile development environment. But, their work requires further elaboration and formalization of requirements of agile SPM process (Vlaanderen et al., 2009).

Harsimarjeet et al. (2011) present PEOR model, which focused on how team members should function to improve organization performance in continuously changing situations. This model minimizes the problem of overtime by continuously monitoring the performance of team members and thus increasing customer's satisfaction. The limitation of this model is that it is proposed for an environment where workers are not permanent (Harsimarjeet et al., 2011).

Nishijima (2013), presents the applicability of agile methodology specially Scrum in traditional development environment. The main problem is cultural resistance to change within organization. This work encourages using agile methodology rather than the traditional, but the success will depend on condition of cultural change and strategy within organization. The limitation of this work is that the client must be committed to the project, has necessary knowledge and be available to answer questions when needed (Nishijima, 2013).

According to Farid et al. (2013), the agile project management methodology has not adequately addressed planning but prioritized activities and nonfunctional requirements. The project management requires suitable quality metrics that would be used to design a risk driven algorithm to prioritize, plan and improve implementation sequence. Non-functional Requirements Planning (NORPLAN) proposes two additional prioritization schemes (Riskiest-Requirement First and Riskiest-Requirement Last). However, there is a need to incorporate other required quality metrics and validate NORPLAN in real world agile software requirements planning teams (Farid et al., 2013).

Guang-yong (2011) presents the software engineering research and practices to improve software quality and productivity. During the process, the project team developing vehicle spare parts management system, import Scrum agile software development using visual studio 2010 as the Scrum process management template. Scrum implementation increases team productivity and quality. But we must explore the continuous expansion and improvement in Scrum, particularly in the area of performance evaluation for in-depth research. In addition, we must examine how to further the integration between different roles (Guang-yong, 2011).

Akif (2012) determines the challenges and issues in Scrum implementation. Issues identified from the survey includes: quality items pileup, module integration issues, code quality, disruption in team work, backlog management, multiple teams, metrics, no technical practices, risk management, mature vs. immature Scrum, sprint duration, release process, lack of Scrum training, documentation, too idealistic Scrum and communication/Scrum ceremonies (Akif , 2012).

**3.    Problem Statement**

The agile methodology is an optimum idea to implement a product. The Scrum methodology is one of the popular methodologies used in organizations. But the main problem in the Scrum is the team size. The Scrum team is usually between 7-9 members and most organizations want to create big projects and need a larger team. However, in a single project large size team may have many issues. This paper attempts to address the problem of fixing the issues in setting up large size team using Scrum.

**4.    The Proposed Solution**
*A.  Overview of Scrum*

The Scrum is an agile methodology to produce a high quality product in short time. There are two important concepts in Scrum; the product backlog and sprint. The product backlog represents the set of requirements for the product and sprint represents an iteration in which a set of activities must be done (Mundra et al., 2013).

### B. Overview of Scrum

There are three core roles in Scrum: PO, SM, and the ST. PO is a key stakeholder of the project and is responsible to visualize and prioritize the features list for the product The SM ensures that the Scrum process is going as agreed, and prevent any impediments that can be faced by the teams e.g. communication, dependencies etc. The ST is responsible for the delivering the product at the end of sprint.

### C. Large Size Team

A typical Scrum team has 7 members plus or minus 2. Software development projects usually are big projects and therefore require a bigger group effort. According to Scrum, large projects need large team to achieve the goal in short time. But large size team in Scrum has some issues.

- Holding a daily meeting with a large size team comprising of 100 people, for example, is impractical.

- Planning and control gets too complicated with large size team.

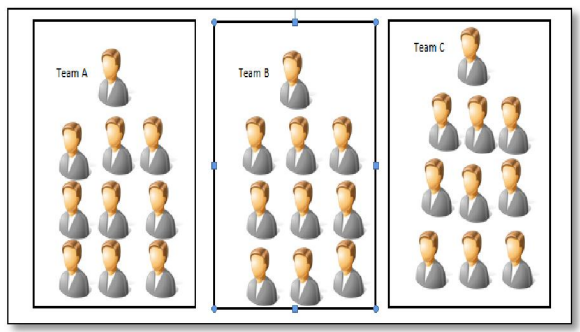

Figure 1. Three teams from team of 30 members

In Scrum of Scrums, we basically divide the team into 2 or more teams, respecting the limit of 7-9 people per team; the Scrum of Scrums team size also depends on the number of teams participating. To understand the Scrum of Scrums concept let us assume this example. We have a project with a team of 30 people. We divide the team into 3 teams each having 10 people.

Considering that all these groups will be working on the same product, there may arise some problems, for example:

- Duplication of work (two teams may implement the same part of the scope).

- Communication failure.

- Integration of different parts of a product developed by different teams.

- Dependencies between tasks of different teams.

The technique that is used to solve these problems is Scrum of Scrums meeting (Paasivaara, 2012).

### A. Scrum of Scrums Meeting

Scrum of Scrums meetings can be consistent through an even higher level meeting called a Scrum of Scrum of Scrums. The Scrum of Scrums meeting is an important technique in scaling Scrum to large project teams. These meetings allow groups of teams to discuss their work, focusing on areas of overlap and integration.

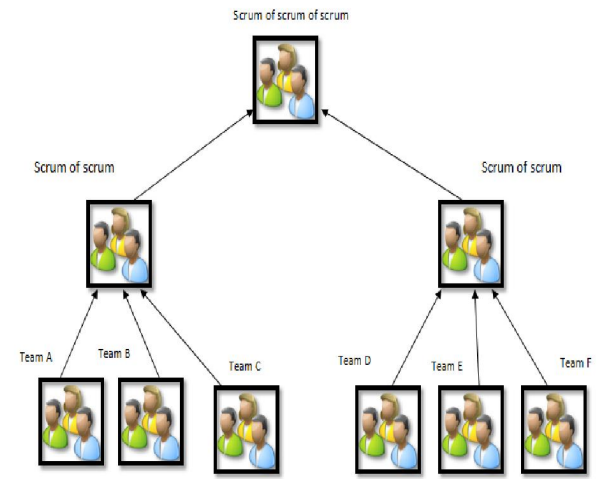

Master. This group then represents the ideal Scrum of Scrums team size. In case of small number of teams participating in the meeting, the teams may choose 2 representatives; a technical contributor, as described above, and a Scrum Master. The attendees of the meeting should change over the course of a project. The team should choose its representative based on who will be in the best position to understand and discuss the issues that arise at that time during a project. For example, early in a project, the issues that are raised at the Scrum of Scrums meeting may focus on technical issues or user experience design. Teams must send a person strong in one of these areas. Later, if the issues are raised around how to collaborate on testing, the tester must be the person chosen for the meeting (Paasivaara, 2012).

*B. Difference between Daily Scrum Meeting and Scrum of Scrums Meeting*

The daily Scrum meeting is not used as a problem-solving or issue resolution meeting. Issues that are raised are usually dealt with by the relevant subgroup immediately after the meeting. In the daily Scrum meeting, each member must answer these three questions:

- What did you do yesterday and today?
- What will you do today?
- Is there any impediment in your way?

In Scrum of Scrums meeting one person, representing his or her entire team, is asked following four questions:

- What has your team done since we last met?
- What will your team do before we meet again?
- Is anything slowing the team down or getting in their way?
- Are you about to put something in another team's way?

*C. Frequency for Scrum of Scrums meeting*

The frequency for Scrum of Scrums meetings should be determined by the team, depending on the complexity and size of project. To make frequency of Scrum meetings easier to understand, let us assume that we have a project with 20 teams each having 5 people. These teams have to be grouped based on some criteria. Let say there are 4 groups of 5 teams. Now we have 20 daily Scrum meetings, 4 daily Scrum group meetings and 1 daily Scrum project meeting. Since the whole project will be divided among 20 teams, so each team will be working on some features of different components of the project. After each team finishes its work, we may face the problem of integration and configuration management. Another consideration in such scenarios is testing of integrated components as a system. The components may work fine when executed standalone, but may not yield when integrated with other components in the whole system.

Scrum of Scrums is a generic model that can be applied to any project, program and portfolio, depending upon the need of the organization; however, the proposed solution has few limitations with respect to team structure and cross-team interaction.

- Team structure limitation: We may divide and organize teams into three levels: upper level, middle level and lower level. This structure may work in some project but may not work in others. It may be more effective to divide teams on the basis of project features.
- Cross-team interaction limitation: The proposed methodology lacks in effective mechanisms to deal with situations when there are multiple product owners or Scrum masters. A particular case of this situation is when some members are working in multiple teams.

**6.    Validation**

First of all, we established main goals to validate our proposed approach. These goals are as under.

Goal 1: Optimize communication between teams in Scrum of Scrums.

Goal 2: Make the system work after integrating all parts.

Goal 3: Reduce the dependencies between the parts of system.

Goal 4: Prevent the duplication of work.

For each goal, a set of meaningful questions was developed that characterized it. Altogether, a questionnaire consisting of 20 questions was distributed among IT professionals. The results were gathered, analysed and scaled using Likert scale (Table 1).

Table 1. Likert scale

| Very low | Low | Nominal | High | Very high |
|---|---|---|---|---|
| 1 | 2 | 3 | 4 | 5 |

Following sections describe, in detail, the findings of this survey.

*A. Cumulative Analysis of Responses to Questions for Goal 1*

Table 2. Participants' response to individual questions for goal 1

| Question # | Cumulative response to questions (%) | | | | |
|---|---|---|---|---|---|
| | very low | low | nominal | high | very high |
| 1 | 20.0 | 32.5 | 37.5 | 10.0 | 0.0 |
| 2 | 0.0 | 0.0 | 32.5 | 50.0 | 17.5 |
| 3 | 0.0 | 2.5 | 10.0 | 35.0 | 52.5 |
| 4 | 0.0 | 2.5 | 7.5 | 40.0 | 50.0 |
| 5 | 2.5 | 2.5 | 7.5 | 37.5 | 50.0 |
| Total | 22.5 | 40.0 | 95.5 | 172.0 | 170 |
| Average | 4.5 | 8.0 | 19.0 | 34.5 | 34.0 |

The table 2 presents participants' response to individual questions for goal 1. We found that 34% of the participants responded as very high and 34.5% responded high, whereas for only 8% the chances of optimum communication were low. Figure 3, depicts cumulative response to all questions for goal 1.

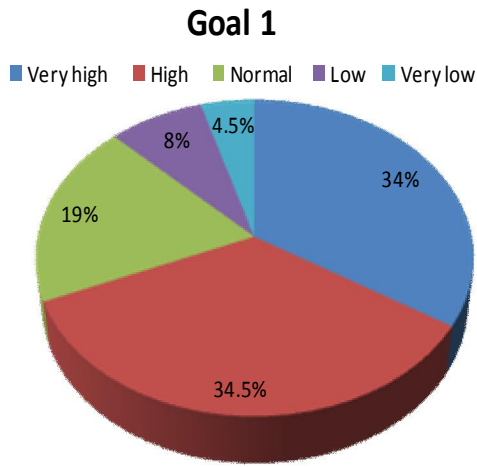

Figure 3. Average response to questions for goal 1

*B.   Cumulative Analysis of Responses to Questions for Goal 2*

The response to survey questions corresponding to goal 2 is shown in Table 3 and Figure 4. It is clear from the analysis of response to our survey questions 6–10 that the likelihood of making system work after integrating all parts is high (57%) whereas for only few the likelihood is low (7%) (Figure 4).

Table 3.   Participants' response to individual questions for goal 2

| Question # | Cumulative response to questions (%) | | | | |
|---|---|---|---|---|---|
| | very low | low | nominal | high | very high |
| 6 | 2.5 | 10.0 | 32.50 | 50.00 | 5.00 |
| 7 | 0.00 | 10.0 | 27.50 | 62.50 | 0.00 |
| 8 | 0.00 | 5.00 | 15.00 | 62.50 | 17.50 |
| 9 | 0.00 | 7.50 | 25.00 | 60.00 | 7.50 |
| 10 | 0.00 | 2.50 | 15.00 | 50.00 | 32.50 |
| Total | 2.5 | 35.0 | 115.0 | 285.0 | 62.5 |
| Average | 0.5 | 7.0 | 23.0 | 57.0 | 12.5 |

*C.   Cumulative Analysis of Responses to Questions for Goal 3*

Table 4 illustrates that questions corresponding to goal 3 retrieved 29.5% answers in agreement with high chances of reducing dependencies between the parts of system. However, 26% and 17.5% of the responses appeared to be low and very low, respectively, as shown below in Figure5

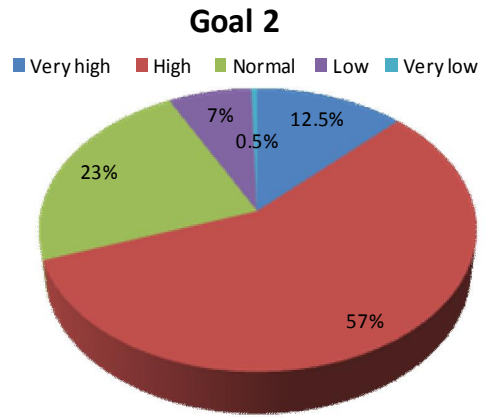

Figure 4. Average response to questions for goal 2

Table 4.   Participants' response to individual questions for goal 3

| Question # | Cumulative response to questions (%) | | | | |
|---|---|---|---|---|---|
| | very low | low | nominal | high | very high |
| 11 | 0.0 | 0.0 | 0.0 | 80.0 | 20.0 |
| 12 | 5.0 | 12.5 | 20.0 | 47.5 | 15.0 |
| 13 | 27.5 | 32.5 | 35.0 | 5.0 | 0.0 |
| 14 | 32.5 | 35.0 | 22.5 | 10.0 | 0.0 |
| 15 | 22.5 | 50.0 | 22.50 | 5.0 | 0.0 |
| Total | 87.5 | 130.0 | 100.0 | 147.5 | 35.0 |
| Average | 17.5 | 26.0 | 20.0 | 29.5 | 7.0 |

*D.   Cumulative Analysis of Responses to Questions for Goal 4*

The result of survey questions for goal 4 are shown in Table 5. The cumulative results of questions (16-20) indicated that 32.5% of the It professionals were in strong agreement (responded very high) with goal 4, i.e. following the proposed solution, the system will work after integration of all parts. Another 24.5% responded as high, 9.5% as very low, 13% as low and 20.5% as nominal. These results are summarized in Figure 6.

*E.   Cumulative Analysis of Responses to All Questions for all Goals*

The average responses of questions for each goal are shown in Table 6. The average of all the goals depicts the overall participants' response to the proposed approach. We found that 57.9% of participants supported the proposed solution with high (36.4%) and very high (21.5%) response.

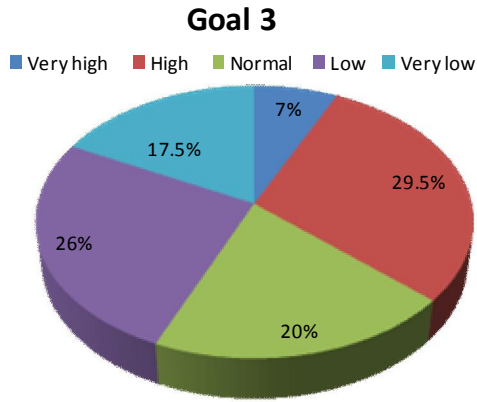

Figure 5. Average response to questions for goal 3

However, 20.6% remained neutral (responded as nominal) and 21.5% responded as low or very low.

Table 5. Participants' response to individual questions for goal 4

| Question # | Cumulative response to questions (%) | | | | |
|---|---|---|---|---|---|
| | very low | low | nominal | high | very high |
| 16 | 12.5 | 25.0 | 35.0 | 27.5 | 0.0 |
| 17 | 2.5 | 5.0 | 20.0 | 50.0 | 22.5 |
| 18 | 32.5 | 32.5 | 10.0 | 17.5 | 7.5 |
| 19 | 0.0 | 0.0 | 25.5 | 17.5 | 57.5 |
| 20 | 0.0 | 2.5 | 12.5 | 10.0 | 75.0 |
| Total | 47.5 | 65.0 | 102.5 | 122.5 | 162.5 |
| Average | 9.5 | 13.0 | 20.5 | 24.5 | 32.5 |

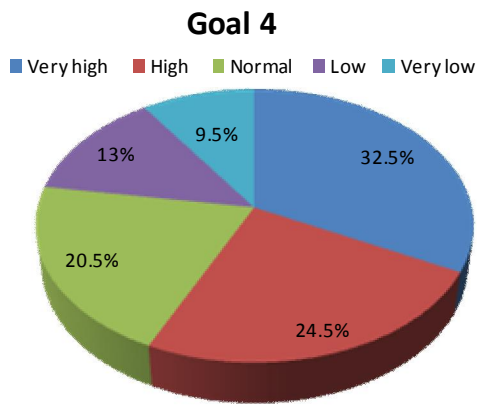

Figure 6. Average response to questions for goal 4

## 7.    Conclusion And Future Work

The computer software domain must keep pace with changes in the environment. Scrum methodology is excellent choice when we want to reduce the development time. Most of the software projects are large and thus require large teams.

Unfortunately, the large size team is the problem in classical Scrum methodology. Large teams can lead to several serious issues including duplication of work, communication failure, and integration of different parts and complex dependencies between tasks done in different teams. We proposed a solution called Scrum of Scrums to address these issues. In Scrum of Scrums, we divided large Scrum teams into teams of the right size, and organize them hierarchically into a Scrum of Scrums. We evaluated the proposed methodology through a survey (conducted with IT professionals). The results show that the majority of the respondents are agreed with the effectiveness of our proposed approach.

Table 6. Overall participants' response to our questions for all goals

| Goals | Overall response to all goals | | | | |
|---|---|---|---|---|---|
| | very low | low | nominal | high | very high |
| 1 | 4.5 | 8.0 | 19.0 | 34.5 | 34.0 |
| 2 | 0.5 | 7.0 | 23.0 | 57.0 | 12.5 |
| 3 | 17.5 | 26.0 | 20.0 | 29.5 | 7.0 |
| 4 | 9.5 | 13.0 | 20.5 | 24.5 | 32.5 |
| Total | 32.0 | 54.0 | 82.5 | 145.5 | 86.0 |
| Average | 8.0 | 13.5 | 20.6 | 36.4 | 21.5 |

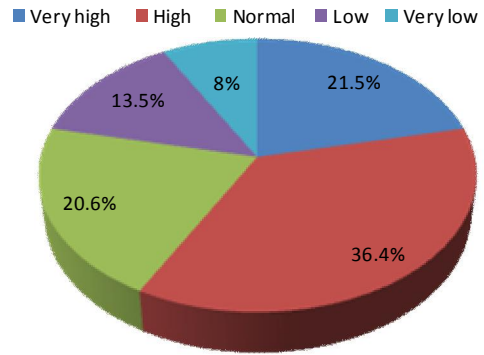

Figure 7. Cumulative response to our questions for all goals

However, the proposed solution has few limitations with respect to team structure and cross-team interaction. We plan to address these limitations in our future work.

5/6/2014